# Comment on
## "Brilliant source of 19.2 attosecond soft X-ray pulses below the atomic unit of time"

Meng Han[1]

*J. R. Macdonald Laboratory, Kansas State University, Manhattan, Kansas, 66506, USA*

Email: meng9@ksu.edu

(Date: October 10, 2025)

Abstract: A recent preprint by Ardana-Lamas *et al*. (Ref. [1], arXiv:2510.04086) re-analyzes a data set of the attosecond streaking experiment on krypton atoms originally reported in Phys Rev X 7, 041030 (2017) [2], and claims the characterization of a 19.2 attosecond light pulse with an overall photon flux of $4.8\times10^{10}$ photons per second. In this comment, we highlight a series of physical and technical issues concerning both the original experiment [2] and the new characterization [1]. Specifically, without accounting for the contribution of Auger electrons or employing filters to remove low-energy harmonics and compensate for the intrinsic pulse chirp, the authors attribute the measured photoelectrons predominantly to the 3d inner-shell orbital of krypton and claim to have retrieved a nearly chirp-free pulse based on the single atomic orbital. In comparison with our recent attosecond streaking experiments on krypton atoms, these physical and technical issues raise significant doubts about the validity of the characterization results.

Characterizing attosecond pulses is crucial for maintaining scientific rigor [3-5]. Attosecond streaking technique is a method for characterization of isolated attosecond light pulses, in which the measured photoelectron traces need to be carefully analyzed to accurately determine the pulse duration [6]. As the spectral bandwidth of attosecond pulses continues to increase, both the streaking measurements and the subsequent pulse retrieval become increasingly challenging [7,8]. As demonstrated in Fig. 1 of a theoretical article [6], which analyzed the experimental data from [2], the theory streaking trace of a 20 as transformed limited pulse differs only very slightly from those of chirped pulses with durations of 90 or 177 attoseconds and the autocorrelation representation shows clearer distinction. Ref. [1] re-analyzed a data set of the attosecond streaking experiment on krypton atoms originally reported in [2], and claims the generation and characterization of a 19.2 attosecond light pulse with an overall photon flux of $4.8\times10^{10}$ photons per second. In the following, we express and detail five fundamental concerns.

**1. Photoelectrons from krypton valence orbitals (4s and 4p) ionized by low-energy harmonics**: In their experiment, they did not measure the harmonic spectrum below 150 eV due to the limited working range of the grating (Hitachi, 2,400 lines per mm according to [9]) and employed no filters to suppress these low-energy components. However, harmonics below 100 eV can generate a significant number of photoelectrons through ionization of the valence orbitals, which would spectrally overlap with the photoelectrons from the 3d inner-shell orbital ionized by higher-energy harmonics (>100 eV). As shown in Fig. 1 of [2], the photoionization cross section of the krypton valence orbitals is substantially larger than that of the 3d orbital in this low-energy energy range. Therefore, attributing the measured photoelectrons solely to ionization of the 3d orbital by harmonics above 100 eV—without showing the HHG spectrum and accounting for their contribution in retrieval—is highly problematic and undermines the reliability of their pulse retrieval.

**2. Intrinsic positive chirp of the attosecond pulses generated by HHG**: Without employing any filters, Ref. [1] reports a nearly chirp-free pulse in Fig. 3b, without specifying the sign of the small residual chirp. However, this result is inconsistent with the combined contribution of the intrinsic attochirp—estimated by the authors themselves to range from 1654 as² to 6172 as²—and the propagation-induced chirp of –101.3 as² predicted by their fluid dynamics simulation (see Fig. 5g), which is only very briefly introduced. A more robust and convincing approach would be to use spectral filters and directly measure the filter-induced chirp from the corresponding streaking traces.

**3. Discrepancy between measured and retrieved streaking traces**: The measured streaking trace shown in Fig. 2c of [1] clearly exhibits sub-cycle asymmetries, which are indicative of a chirped attosecond pulse. However, the retrieved trace presented in Fig. 2b of [1] appears artificially "perfect," lacking any such asymmetry. This discrepancy between the measured and retrieved streaking traces raises serious concerns about the reliability of their pulse characterization. Moreover, the streaking trace in Fig. 2c of [1] appears very similar to Fig. 4 of [2], which was later retrieved using the autocorrelation representation in [6] to yield a pulse duration of ~165 as. In contrast, the current retrieval result is about eight times shorter, raising questions about how such a significant discrepancy arises.

**4. Non-negligible contribution of Auger electrons for atomic inner-shell ionization:** We have also conducted attosecond streaking experiments on krypton atoms [10] using isolated attosecond pulses with a central photon energy of ~130 eV (bandwidth ~20 eV), where inner-shell ionization resulted in a significant yield of Auger electrons. These electrons are notably absent in both the experimental data and retrieval process presented in [1]. As shown in Figure 1, in the energy range of 30–60 eV the Auger electron yield is comparable to that of the 3d photoelectrons, agreeing with the measurement with synchrotrons [11]. Importantly, this electron energy range was used for retrieving the attosecond pulse in Fig. 3b of [1]. Ignoring the contribution of Auger electrons in the pulse retrieval process raises serious concerns about the accuracy and validity of their characterization. Additionally, Refs. [1,2] did not show the energy calibration of the time-of-flight (TOF) electron spectrometer—a crucial step, particularly for high-energy electrons [7]. This calibration is typically performed using the energies of Auger electrons.

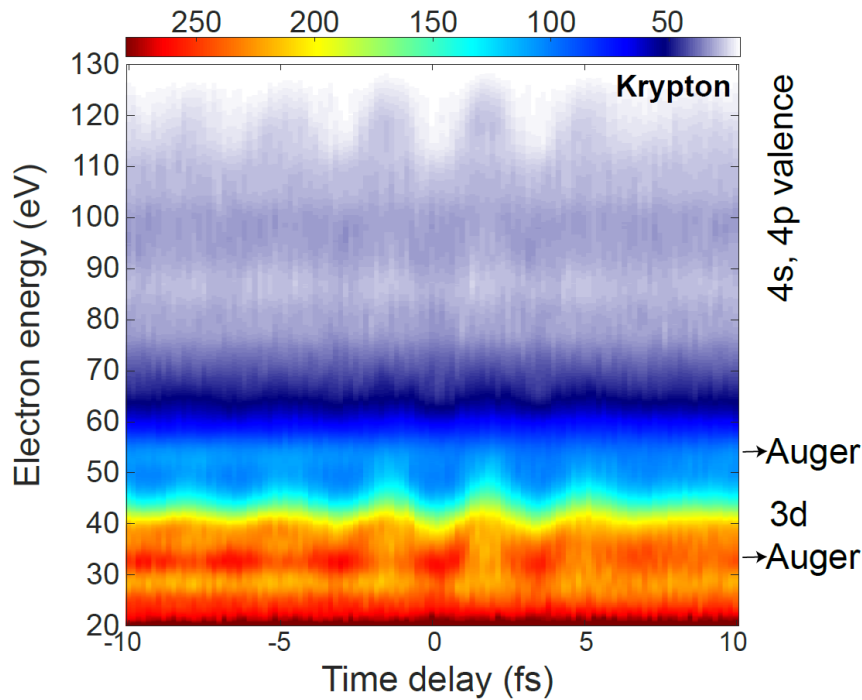

Figure 1. Attosecond streaking experiment on krypton atoms [10] by an attosecond soft-x-ray pulse centered at 130 eV after a 200-nm-thick silver filter. Two Auger electron bands are observed and overlapped with the 3d photoelectrons.

**5. Photon flux calibration and discrepancy**: Ref. [1] claims an overall photon flux of $4.8\times10^{10}$ photons per second, but provides no measurement details regarding the calibration procedure. Accurate photon flux measurement requires the use of a filter to fully remove the residual driving laser field. However, as mentioned above, that work did not use any filter to remove driving field. Notably, the reported photon flux in [1] is almost higher by five orders of magnitude compared to the previously published value of $5.6\times10^{5}$ photons per second from 284 to 350 eV on target under the similar experimental conditions in [2].

In summary, the krypton atom is not a suitable ionization target for streaking experiments aimed at characterizing super-broadband attosecond soft X-ray pulses. In Ref. [1], the data analysis did not provide a convincing pulse characterization.

**References**

[1] Fernando Ardana-Lamas, Seth L. Cousin, Juliette Lignieres, Jens Biegert, Brilliant source of 19.2 attosecond soft X-ray pulses below the atomic unit of time, arXiv:2510.04086 (2025). https://doi.org/10.48550/arXiv.2510.04086

[2] S. L. Cousin, N. Di Palo, B. Buades, S. M. Teichmann, M. Reduzzi, M. Devetta, A. Kheifets, G. Sansone, J. Biegert, Attosecond Streaking in the Water Window: A New Regime of Attosecond Pulse Characterization. Phys Rev X 7, 041030 (2017).

[3] G. Sansone, E. Benedetti, F. Calegari, C. Vozzi, L. Avaldi, R. Flammini, L. Poletto, P. Villoresi, C. Altucci, R. Velotta, S. Stagira, S. De Silvestri, M. Nisoli, Isolated Single-Cycle Attosecond Pulses. Science 314, 443–446 (2006).

[4] M. Chini, K. Zhao, Z. Chang. The generation, characterization and applications of broadband isolated attosecond pulses. Nature Photonics, 8(3): 178-186 (2014).

[5] Katsumi Midorikawa. Progress on table-top isolated attosecond light sources. Nature Photonics, 16(4):267–278, (2022).

[6] X. Zhao, S. Wang, W. Yu, H. Wei, C. Wei, B. Wang, J. Chen, and C. D. Lin, Metrology of Time-Domain Soft X-Ray Attosecond Pulses and Reevaluation of Pulse Durations of Three Recent Experiments. Phys. Rev. Applied 13, 034043 (2020).

[7] T. Gaumnitz, A. Jain, Y. Pertot, M. Huppert, I. Jordan, F. Ardana-Lamas, and H. J. Worner. Streaking of 43-attosecond soft-x-ray pulses generated by a passively cep-stable mid-infrared driver. Optics express, 25(22):27506–27518, (2017).

[8] J. Li, X. Ren, Y. Yin, K. Zhao, A. Chew, Y. Cheng, E. Cunningham, Y. Wang, S. Hu, Y. Wu, M. Chini, Z. Chang. 53-attosecond x-ray pulses reach the carbon k-edge. Nature communications, 8(1):186, (2017).

[9] S. M. Teichmann, F. Silva, S. L. Cousin, M. Hemmer and J. Biegert, 0.5-keV Soft X-ray attosecond continua. Nat Commun 7, 11493 (2016).

[10] M. Hasan, J. Gao, H. Liang *et al*, 2025, (to be submitted).

[11] B. Schmidtke, T. Khalil, M. Drescher, N. Müller, N. Kabachnik and U. Heinzmann, The Kr $M_{4,5}N_1N_{2,3}$ $^1P_1$ Auger decay: measurement of the transferred spin polarization and analysis of Auger amplitudes. Journal of Physics B: Atomic, Molecular and Optical Physics, 34, 4293 (2001).